\documentclass{elsart}
\usepackage{graphicx}
\usepackage{epsfig}

\begin{document}
\begin{frontmatter}
\title{The few scales of nuclei and nuclear matter}
\author[UFF]{A. Delfino},
\author[ITA]{T. Frederico},
 \author[CESET]{V. S. Tim\'oteo}, and
\author[IFT]{Lauro Tomio}
\address[UFF]
{Instituto de F\'\i sica, Universidade Federal Fluminense,
24210-900 Niter\'oi, RJ, Brasil}
\address[ITA]
{Departamento de F\'\i sica, Instituto Tecnol\'ogico de
Aeron\'autica, CTA, 12228-900, S\~ao Jos\'e dos Campos, Brasil}
\address[CESET]{Centro Superior de Educa\c c\~ao Tecnol\'ogica,
Universidade Estadual de Campinas, 13484-370, Limeira, SP, Brasil}
  \address[IFT]{Instituto de F\'\i sica Te\'orica, Universidade
Estadual Paulista,  01405-900, S\~{a}o Paulo, Brasil }
\date{\today}

\begin{abstract}
The well known correlations of low-energy three and four-nucleon
observables with a typical three-nucleon scale (e.g. the Tjon line)
is extended to light nuclei and nuclear matter.
Evidence for the scaling between light nuclei binding energies
and the triton one are  pointed out.
We show that the saturation energy and density of nuclear matter
are correlated to the triton binding. From the available systematic
nuclear matter calculations, we verify the existence of bands
representing these correlations.
\end{abstract}
\maketitle
\end{frontmatter}

Two-nucleon interactions are typically constructed to fit
scattering data and deuteron properties. When such interactions
are used to calculate three-nucleon observables, the results
exhibit some discrepancies~\cite{gloeckle}. Basically, they are
explained as originated from different strengths of the
two-nucleon tensor force and short-range repulsions, provided that
all realistic two-nucleon interactions have the correct one-pion
exchange tail. In four-nucleon bound state ($^4$He ) calculations
the discrepancies still remain. But, at least, they are
correlated, as seen in the  binding energies of $^4$He
($B_\alpha$) and triton ($B_t$), which lie on a very narrow
band~\cite{Tjon}, obtained when the short-range repulsion of the
nucleon-nucleon interaction is varied while two-nucleon
informations (deuteron and scattering) are kept fixed. This
correlation is known as Tjon line~\cite{Tjon}. $B_\alpha$ and
$B_t$ follows an almost straight line in the range of about 1-2 MeV 
of variation of the triton binding energy around the
experimental value. As the long-range two-nucleon scales we have
the deuteron binding energy ($B_d$) and the singlet virtual-state
energy ($B_v$).

Two-body short-ranged interactions, supporting very low two-body
binding energy and/or large scattering lengths, when used to
calculate three-body systems, approach what we call the
universal Thomas-Efimov limit~\cite{adhikari}.  By trying to find
the range $r_0$ of the two-nucleon force, Thomas~\cite{thomas}
showed that when $r_0 \rightarrow 0$, while the two-body binding
energy $B_2$ is kept fixed, the three-body binding energy goes to
infinity (Thomas collapse). Much latter, Efimov~\cite{Efimov}
showed that, in the limit $B_2 = 0 (r_0 \not = 0)$ the number of
three-body bound states is infinite with an accumulation point at
the common two- and three-body threshold. Note that both the
Thomas and Efimov effects are claimed to be model independent,
since they are due to a dynamically generated effective three-body
potential acting at distances outside the range of the two-body
potential. These apparently different effects are related to the
same scaling mechanism, as shown in Ref.~\cite{adhikari}. { In
other words, the Thomas effect appears when $r_0$ is much smaller
than the size of the two-body system (which is of the order of the
scattering length $|a|$), while the Efimov effect arises for
$|a|>>r_0$. Therefore, what matters for both
effects is the same condition: $|a|>>r_0$ or the ratio
$|a|/r_0>>1$. In terms of the two-body energies this is
translated to $\sqrt{\left(m|B_2|/\hbar^2\right)}\;r_0<<1$ 
($m$ the boson mass).}
One would expect that the Thomas-Efimov effect is manifested in
weakly-bound quantum few-body systems which are much larger in
size than the corresponding two-body effective range.
Notice that a zero binding energy for a free two-body system is
not known in nature. But, nowadays it was shown that, for trapped
ultracold gases of certain atomic species, it is possible to
adjust the two-body scattering length at very large values, using
Feshbach resonance techniques, by tuning the external magnetic
field~\cite{roberts}. In this case, it is expected that the
Thomas-Efimov effect can be manifested~\cite{jonsell}.

The deuteron and triton may be viewed as low energy systems with
large size scales in which the range of the potential is smaller
than the corresponding healing distances of the wave functions,
leading the nucleons to have a high probability to be outside of
the interaction range. Then, the low-energy properties of these
systems can be studied with models that minimally includes the
physics of the Thomas-Efimov effect, as in the case of a few-body
model with renormalized pairwise s-wave zero-range
force~\cite{ren}. This approach shows that all the low-energy
properties of the three-body system are well defined in the model,
once one three-body scale and the two-body low-energy observables
are given. As a consequence, correlations between two three-body
$s-$wave observables are expected to appear in model calculations
with short-ranged interactions.
Along this line, some previous works (see \cite{few} and references
therein) have studied weakly-bound halo states in exotic nuclei as
well as possible Efimov states for He-He-Alkali molecules.

The scaling of three-nucleon observables with the triton binding
energy corresponds to universal behaviors found when a three-body
scale is varied. For example, the Phillips plot~\cite{phillips} of
the neutron-deuteron doublet scattering length, as a function of
the triton energy is nowadays one of the universal scalings found
in the three-nucleon system~\cite{pol88,bed}. In general, it is
observed for nuclear and molecular weakly bound three-body
systems~\cite{prepjensen,rvmpjensen,braaten} the scaling of
observables with the three-body binding energy.

The Thomas-collapse of the three-body energy in systems of maximum
wave-function symmetry implies the existence of a three nucleon
scale (identified with the triton binding energy)
governing the short-range behavior of the wave-function.
Four nucleons can also form a state of maximum symmetry, allowing
in principle the collapse of such configuration, independently of
the three-nucleon collapse~\cite{ren}.
This is under discussion and it is suggested in \cite{platter4b}
that the four-body scale is not independent of the three-body one.
However, in their work it was introduced a three-body force to
stabilize the shallowest three-body state, against the variation
of the cut-off. The three-body interaction can be attractive or
repulsive and their conclusion lies on the repulsive sector. We
note that the attractive part indicates a possible independent
behavior of the four body ground-state energy from the three-body
one. Certainly this point merits further discussions and so far,
we think, it is still open the possibility of a four-body scale.
Anyway, as the nucleon-nucleon interaction is strongly repulsive
at short range and therefore the probability of four nucleons to
be simultaneously in a volume $\sim r_0^3$ is quite small,
presumably the four-nucleon scale itself has much less opportunity
to be evidenced in realistic nuclear models.
Indeed, this is indicated by the existence of the Tjon line. Due
to that, as we will see later, the four-nucleon binding energy is
eliminated in favor of the triton binding energy. { In this
respect it is worthwhile to note that Platter et al. extended the
effective field theory framework applied to
four-bosons~\cite{platter4b} to calculate the $^4$He binding
energy by controlling the triton energy through a repulsive
effective three-nucleon force. Within their
approach~\cite{platter2005} the Tjon line is reproduced.}

In a nuclear scenario dominated by an interaction with a range
smaller than the nucleon-nucleon scattering lengths, and
considering the triton and $^4$He nuclear sizes yet larger than
the force range, the picture of nuclei would be of a many-body
system with the wave-function being an eigenfunction of the free
Hamiltonian almost everywhere.
The Pauli principle allows only up to four nucleons at the same
position, forbidding certain particular configurations with
overlap of more particles. If more than four particles are allowed
to overlap, it would imply  that the asymptotic information from
the interaction of the cluster would go beyond of those already
fixed by the low-energy observables of two, three and four
nucleons. By some unknown reason the parameters of Quantum Chromodynamics
are close to this limit. {It was conjectured in Ref.~\cite{braatenprl} 
that a small change in the light quark masses away from their 
physical values} could put the deuteron and the
singlet virtual state at zero binding energy, and therefore the
above idealized picture of the nuclear systems could not be far
from reality. It is quite amazing thinking that nuclear wave
functions could heal much beyond the interaction range. Therefore,
the details of the long wavelength structure of nuclei are given
by the free Hamiltonian and by few-nucleon scales, which determine
the wave function at short distances. The universal behavior of
the scaling functions are due to that.

If one wonders about the neutron matter within a non-relativistic
quantum framework, in the limit of a zero-range force, we could
say that the only scale in this case is the neutron-neutron
scattering length. Therefore, the binding energy of neutron
droplets will be strongly correlated to that quantity, which is
the only physical scale in this situation allowed by the Pauli
principle. This discussion has been performed in the context of
three neutron systems~\cite{smatrix}. Moreover, it was concluded
in Ref.~\cite{piepern} that stable tetra-neutron droplets would
imply a major change in the neutron-neutron scattering length.

Another example of the dominance of only two-body scale appears in
three-boson systems in two dimensions, where the Thomas-Efimov
effect is absent~\cite{adhikari}. In this case, only two-body low energy scales
are enough to define the many-body properties in the limit of a
zero-range interaction. The low-energy properties of a many-body
system of spin-zero particles in two dimensions will be sensitive
only to the two-boson binding energy. Even in the case where
bosons are trapped, since the essential singularity of the
point-like configuration is not affected by the confining force as
the harmonic one.

For the sake of generality, we start with the observables $B_d$,
$B_v$, $B_t$ and $B_\alpha$ as the scales determining the
asymptotic properties of nuclei~\cite{few}. Then, in the limit of
a zero-range interaction, we write the binding energy of a nucleus
with mass number $A$ and isospin projection $I_z$, considering
isospin breaking effects, as
\begin{eqnarray}
B_{(A,~I_z)} \,=\, A~ B_t~\mathcal{B}\left(\beta_v,\beta_d,
 \beta_\alpha,A,I_z\right) ,
\label{tjl}
\end{eqnarray}
where $\beta_a=B_a/B_t$ with $a=$ $v$, $d$ and $\alpha$.

According to the Tjon line, $\beta_{\alpha}$ remains approximately
constant for a variety of two-nucleon potentials and the
parametrization of the numerical results, given in MeV, for several
two-nucleon potentials is
\begin{eqnarray}
B_\alpha = 4.72 ~\left(B_t - 2.48\right) \ , \label{tjon}
\end{eqnarray}
which for $B_t^{exp}=8.48$~MeV gives $B_\alpha^{exp}=28.32$~MeV.
Using (\ref{tjon}) in (\ref{tjl}),
\begin{eqnarray}
R_{(A,~I_z)}=B_{(A,~I_z)}/A =\, ~ B_t~\mathcal{R}\left(B_t,A,I_z
\right) , \label{en0}
\end{eqnarray}
where in the scaling function $\mathcal{R}_{(A,~I_z)}$ the values
of $B_d$ and $B_v$ are fixed to the experimental values. The
dependence of $B_\alpha$ with $B_t$ for realistic nucleon-nucleon
potentials is given by Eq.~(\ref{tjon}).

Equation (\ref{en0}) generalizes the concept of the Tjon line to
nuclei. Recent  calculations using the AV18 nucleon-nucleon
potential plus three-body forces~\cite{pieper} show that there is
a systematic improvement of the binding energy results for He, Li,
Be and B isotopes simultaneously with the triton binding energy,
when models are tuned to fit $B_t$. It is important to note
that these AV18 calculations have at least two three-body
parameters that are fitted to $B_t$ and nuclear matter saturation
properties. Consequently, one could argue that such calculations
cannot provide evidence for one-parameter correlation. The fitting
to nuclear matter calculation presumably is not that important for
light nuclei in view of the dominance of the triton binding (or
three-body correlations) in the four-nucleon bound state as given
by the Tjon line. Therefore, it is reasonable to think  that
three-body correlations are  quite important for light-nuclei,
since that even for the alpha particle where the nucleons are in a
very compact configuration this occurs. In our opinion, the
fitting of nuclear matter saturation properties has more to do
with the approximations done in nuclear matter calculations. The
three-body potential should be somewhat tuned, which is, probably,
not so important for light nuclei once the triton binding attains
its physical value.

For nuclear matter properties calculation using a variety of
two-nucleon potentials, in which the tensor strength was varied
but the deuteron binding energy was kept fixed, it was shown that
these interactions cannot quantitatively account for nuclear
saturation~\cite{coester,day}. Coester et al.~\cite{coester}
observed that, in an energy versus density plot, the saturation
points of nuclear matter obtained by employing different realistic
potentials are located along a band ('Coester band'). Also, in a
relativistic framework it was observed such strong
correlation~\cite{BJP}. The displayed nuclear matter binding
energy ($B_A/A\equiv B_{(A,0)}/A$) versus saturation density
[$\,\rho_{o} = (2/3)k_{F}^{3}/ \pi^{2} $, with $k_F$ the Fermi
momentum] results are within a narrow band~\cite{coester}. The
observation given in Ref.~\cite{coester} have been studied by many
other authors that have used nuclear matter binding energies and
saturation densities from different two-nucleon interactions. The
main argumentation, as also in the case of three-nucleon
calculations, is that this effect comes from different strengths
of the two-nucleon tensor force and short range repulsion, which
changes the triton binding energy, while keeping fixed the
low-energy two-body scales.

\begin{figure}[t]
\centerline{\epsfig{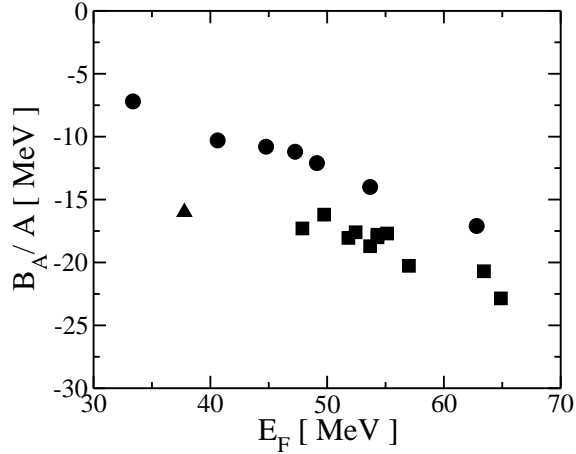}} \vspace{.1cm}
\caption{
Infinite nuclear matter binding energy as a
function of $E_F$ extracted from Ref.~\cite{mac} (solid circles
and squares). The squares includes the single particle
contribution in the continuum. The full triangle is given by the
empirical values.}
 \label{fig1}
\end{figure}

Basically, nuclear matter saturates due to the composed repulsive
and attractive short-range two-nucleon potential. { Since, it
may also be seen as a typical low-energy problem,} it is natural
to question whether any connection exists between the proper
few-body scales, $\,B_{d}\,$, $\,B_{v}\,$ and $\,B_{t}\,$ with
those of the many-body problem, like the $\,B_A/A$ and the Fermi
energy $\,E_F=\hbar^2 k^2_{F}/(2m_N)\,$. For light nuclei there is
strong evidences of scaling between $\,B_{d}\,$, $\,B_{v}\,$ and
$\,B_{t}\,$ as expressed by Eq.~(\ref{en0}).

Here we are arguing that the scales of nuclear matter, $\,B_A/A$
and $E_F$, are determined by $\,B_{d}\,$, $\,B_{v}\,$ and
$\,B_{t}\,$. Therefore, we suppose that going to the
infinite isospin symmetrical nuclear matter, $A\rightarrow \infty$
and $I_z=0$, the limit
\begin{eqnarray}
\frac{B_A}{A}&\doteq& \frac{B_t}{A} \lim_{A\rightarrow
\infty}\mathcal{B}\left(\beta_v,\beta_d,
 \beta_\alpha,A,I_z=0\right)
\nonumber \\ &=& B_t~ \mathcal{G}\left(\beta_v,\beta_d,
 \beta_\alpha\right)
 , \label{esnm}
\end{eqnarray}
is well defined and expresses the correlation between the binding
energy of the nucleon in nuclear matter with the few-nucleon
scales. The Fermi energy
\begin{eqnarray}
E_F \,=\, B_t~\mathcal{E}_F\left(\beta_v,\beta_d,
 \beta_\alpha\right), \label{ef1}
\end{eqnarray}
will be correlated as well to the few-nucleon binding energies.

The aim of this work is to study the possible correlation of the
nuclear matter binding energy per nucleon with $\,B_{d}\,$,
$\,B_{v}\,$ and $\,B_{t}\,$, in order to improve our understanding
of the general and important scaling of observables. Our
investigation is based on Eqs.~(\ref{esnm}) and (\ref{ef1}),
motivated by our previous discussion that leads to Eq.~(\ref{tjl})
and in the several works that recognize the role played by the
low-energy few-body
scales~\cite{ren,few,prepjensen,bed,rvmpjensen,braaten}, in defining
the observables of few-nucleon systems.

In the present framework, the universal scaling functions connect
the proper scales of the few-body system with those of the
many-body system, as given by Eqs.~(\ref{esnm}) and (\ref{ef1}).
Different potentials, which describe the deuteron and the
two-nucleon scattering properties give different values of
$\,B_{t}\,$ , $\,B_{\alpha}\,$  $\,B_{A}/A\,$ and $\,E_{F}\,$.  As
we have done in deriving  Eq.~(\ref{en0}) for a class of changes
in the short-range part of the nuclear force that keeps the
deuteron and low energy scattering properties unchanged, and
taking into account that for these variations of the potential the
$^4$He and triton binding energies are strongly correlated as
given by the Tjon line, one can rewrite Eqs.~(\ref{esnm}) and
(\ref{ef1}) in order to get a one parameter scaling:
\begin{eqnarray}
\frac{B_A}A &=& B_t~\mathcal{G}\left(B_t\right) \label{esnm2}
\end{eqnarray}
for fixed $B_d$ and $B_v$, where the only true dependence in the
class of potential variations is dominated by $B_t$.  The
analogous expression for the Fermi energy is
\begin{eqnarray}
E_F \,=\, B_t~\mathcal{E}_F\left(B_t \right), \label{ef2}
\end{eqnarray}
where $E_F$ scale with $B_t$.

\begin{figure}[t]
\centerline{\epsfig{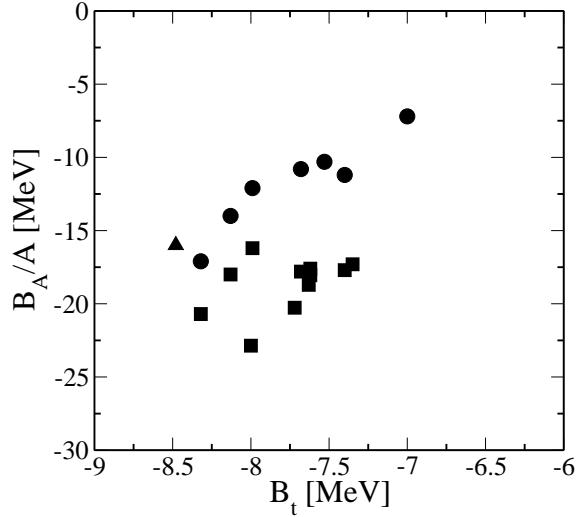}} 
\caption{${B_A}/A$ as a function of $B_t$ extracted from
Ref.~\cite{mac} (solid circles and squares). The squares includes
the single particle contribution in the continuum. The full triangle
is given by the empirical values. } \label{fig2}
\end{figure}

In the perspective of the one parameter functions of
Eqs.~(\ref{esnm2}) and (\ref{ef2}), it is clear that one could
express $E_t$ as a function of $E_F$ and  immediately get
\begin{eqnarray}
\frac{B_{A}}{A~E_{F}}    \,=\, \mathcal{C}( E_F), \label{ratio}
\end{eqnarray}
the correlation implied by the Coester band.

In order to enlighten our discussion we bring a variety of nuclear
matter binding energies $\,B_{A}/A\,$ at the corresponding
saturation density, represented by the Fermi momenta $\,k_{F}\,$,
calculated from different two-nucleon potentials. In Fig.
\ref{fig1}, we present the well known Coester band in which the
results for $\,B_A/A$ and $\,E_{F}\,$ are showed. The two distinct
bands represent the nuclear matter calculations with and without
the single-particle continuum contributions. The empirical values
are $B_A/A=16$~MeV and $E_F=37.8$~MeV from ~\cite{expnm}.

\begin{figure}[t]
\centerline{\epsfig{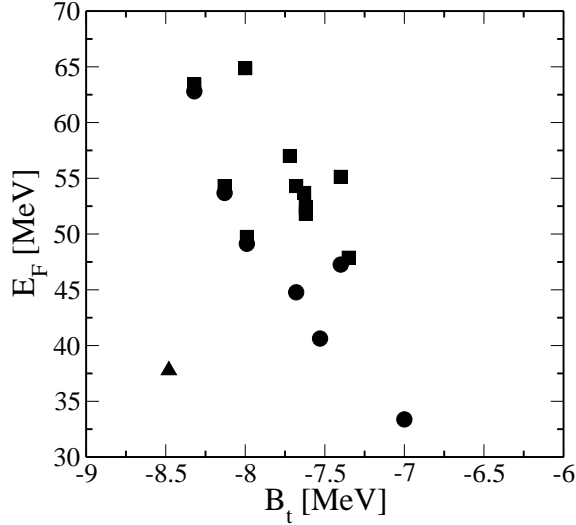}} 
\caption{${E_F}$ as a function of $B_t$  extracted from
Ref.~\cite{mac} (solid circles and squares). The squares includes
the single particle contribution in the continuum. The full triangle
is given by the empirical values.} \label{fig3}
\end{figure}

The correlation of $B_A/A$ with $B_t$ expressed by (\ref{esnm2})
is plotted in Fig.~\ref{fig2}, for the same set of results given
in Fig.~\ref{fig1}. The two-nucleon potentials present different
values for the triton binding energy $\,B_{t}\,$, while the
two-nucleon low-energy observables are fixed. We observe that the
scaling function $E_t{\mathcal G}(E_t)\,$ is quite linear in the
interval of about 2~MeV including the triton binding energy. {
We note that the ratio $B_A/A$ depends strongly on $B_t$. We
understand this fact as a reminiscent manifestation of the
three-body scale in the nuclear matter results obtained with only
two-body correlations.} In Fig.~\ref{fig3}, we show the
correlation between the Fermi energy and the triton binding
energy. In general, we observe that the increase in the three-body
scale leads to the increase of the Fermi energy, which is
reasonable in view of the scaling function (\ref{ef2}). However,
we observe that the empirical values disagrees with the general
trend of the correlation, a problem that could already be
anticipated by looking at the Coester band in Fig.~\ref{fig1}.

The inclusion of three-body correlations, which carries the
dynamics that stabilizes the Thomas collapse, presumably has a
repulsive effect diminishing the saturation density and somewhat
the nuclear matter binding.  Due the short-range repulsion of the
nucleon-nucleon interaction, the nuclear matter tends to saturate
at large densities if only two-body correlations are considered
and the empirical binding is achieved. The dynamically generated
three-body stabilization mechanism carried only through the
three-body correlations should appear in the nucleon-nucleon
interaction range, however such repulsive contribution is absent
if only two-body correlations are considered in the evaluation of
nuclear matter properties. Therefore, we suspect that the inclusion
of three-body correlations in nuclear matter calculations will
bring the correlation curve of the $B_A/A$ with $B_t$ in
Fig.~\ref{fig2} toward the empirical values; and also in
Fig.~\ref{fig3}, where the saturation densities would be possibly
found at lower values for a given triton binding energy.

{Recent sophisticated many-body calculations~\cite{monte},
where the triton binding energy and nuclear matter saturation
density were adjusted through two three-body parameters make
subtle the clear appreciation of the real role of the three-body
correlations, once the fit of these realistic forces mixes
differences that come from the interaction and from the many-body
approximations in nuclear matter calculations. It is beyond our
work the discussion of the quite involved many-body approximations
needed to perform such calculations. Nevertheless, our conjecture
is obviously based on qualitative arguments, which implies that
just one three-body scale parameter is relevant for a systematic
description of light nuclei and nuclear matter.}

\begin{figure}[t]
\centerline{\epsfig{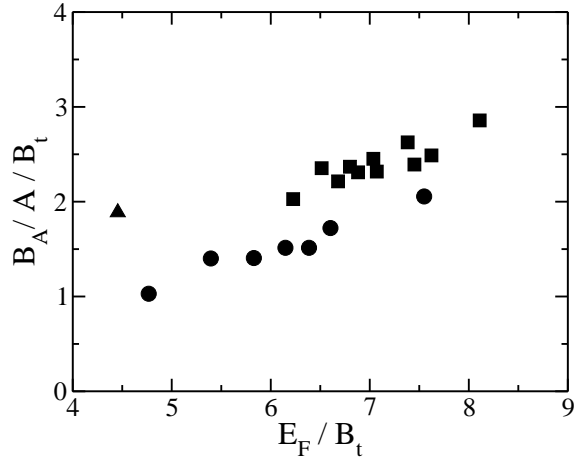}}
 \caption{Infinite nuclear matter binding energy as a function of
$E_F$, both in units of the triton binding energy. The calculation
results are extracted from Ref.~\cite{mac} (solid circles and squares). 
The squares includes the single particle contribution in the continuum.
The full triangle represents the empirical values. } \label{fig4}
\end{figure}

The  one parameter dependence in Eqs.~(\ref{esnm2}) and
(\ref{ef2}) suggests to plot the dimensionless quantity $B_A/(A
B_t)$ as a function of the ratio $E_F/B_t$, which should look as
an almost linear correlation. We display in Fig.~\ref{fig4} the
values for $\,B_{A}/(A~B_{t})\,$ versus $\,E_F/B_{t}\,$.  As we
could anticipate, the results show a clear linear correlation. We
are tempted to say that, if the correlation is extrapolated and
assuming that the binding and saturation densities somewhat
decreases when three-body correlations are considered, it looks to
be possible that the empirical values would be consistent with the
correlation band.

In summary, we suggest for the first time a possible scaling of
nuclei asymptotic properties, in particular the nuclear binding
energies with the triton binding energy, substantiated by recent
realistic calculations of light nuclei. This observation
generalizes to the many-nucleon context the correlations between
observables found in the three and four-nucleon systems. Beyond
that, we found that the original correlation between the nuclear
matter binding energy per nucleon with the Fermi momentum
described by the Coester band can now be seen as robustly
represented by the scaling of nuclear matter properties with the
triton binding energy. The values of $B_t$ carry different aspects
of the used two- and three-nucleon potentials. { To
verify the extension of our conjecture, we propose that one could
control the strength of the three-body force without a many-body
parameter in order produce different triton binding energies and
nuclear matter properties. We emphasizing that the nuclear matter
results should be obtained within the same approximations for all
models. In this way, we expect that the plotted results in figures
2 and 3 should lie in a very narrow band.}

Our discussion may turn in a simple way to systematize results of
possible forthcoming realistic calculations for many-nucleon
systems. Consistent with our conclusions, it was argued in
Ref.~\cite{braatenprl} that QCD implies that nuclear physics is
 close to the Thomas-Efimov limit. { In this reference, it
was conjectured that a small change in the light quark masses away
from their physical values} could be enough to move the deuteron
and the singlet virtual state to zero binding energies. Therefore,
nuclear physics could be dominated by long-range universal
effective forces making the scalings not only a subtlety but also
an evident reality in ab-initio non-relativistic nuclear model
calculations.

{\it Acknowledgments.} This work was partially supported by
Funda\c {c}\~{a}o de Amparo \`{a} Pesquisa do Estado de S\~{a}o
Paulo and Conselho Nacional de Desenvolvimento Cient\'{\i}fico e
Tecnol\'{o}gico. V. S. T. would like to thank FAEPEX/UNICAMP for
partial financial support.

\end{document}